# Charge Transport in a Zn-Porphyrin single molecule junction

Mickael L. Perrin[1], Christian A. Martin[1], Ferry Prins[1], Ahson J. Shaikh[2], Rienk Eelkema[2], Jan H. van Esch[2], Jan M. van Ruitenbeek[3], Herre S. J. van der Zant[1], Diana Dulić*[1]

1 Kavli Institute of Nanoscience, Delft University of Technology, Lorentzweg 1, Delft, The Netherlands

2 Department of Chemical Engineering, Delft University of Technology, Julianalaan 136, 2628 BL Delft, The Netherlands

3 Kamerlingh Onnes Laboratory, Leiden University, Niels Bohrweg 2, 2333 CA Leiden, The Netherlands

* Corresponding author Diana Dulić – d.dulic@tudelft.nl

## Abstract

We have investigated charge transport in ZnTPPdT-Pyr molecular junctions using the lithographic MCBJ technique at room temperature and cryogenic temperature (6K). We combined low-bias statistical measurements with spectroscopy of the molecular levels using I(V) characteristics. This combination allows us to characterize the transport in a molecular junction in detail. This complex molecule can form different junction configurations, which is observed in trace histograms and in current-voltage (I(V)) measurements. Both methods show that multiple stable single–molecule junction configurations can be obtained by modulating the inter-electrode distance. In addition we demonstrate that different ZnTPPdT-Pyr junction configurations can lead to completely different spectroscopic features with the same conductance values. We show that statistical low-bias conductance measurements should be interpreted with care, and that the combination with I(V) spectroscopy represents an essential tool for a more detailed characterization of the charge transport in a single molecule.

# Keywords

Mechanically controllable break junction; molecular configuration; molecular electronics; porphyrin; single-molecule transport

# Introduction

The break junction method represents a popular choice to investigate the electronic transport through metal-molecule-metal junctions [1-6]. While repeatingly breaking and fusing two metallic electrodes, the low-bias conductance is monitored as a function of electrode displacement. Such low bias transport measurements have been extensively used to study the molecular conductance dependence of rod-like molecules on their length [1,2], their conformation [3,4] and their anchoring groups [5,6]. However, as the bias range is very limited, the main contribution to the current is off-resonant transport. As such, spectroscopic information about molecular energy levels involved in the charge transport is lacking.

Here, we investigate charge transport through a Zinc porphyrin with an axial pyridine ligand in both the low-bias and the high-bias regime. Porphyrins are interesting for this purpose as they are complex, non-rod-like molecules that can form different stable configurations [7,8], especially when functionalized with metal-bound axial pyridine ligands [9]. Using the mechanically controllable break junction technique

(MCBJ), we study the low-bias conductance as a function of electrode displacement. In addition, we use current-voltage measurements for different electrode spacings to gain spectroscopic information in the high-bias regime.

The MCBJ technique is an elegant way to control the spacing between two metallic electrodes with a sub-atomic (< $10^{-10}$ m) resolution [10-12]. This control is achieved by bending a substrate with a pair of partially suspended electrodes in a three-point bending mechanism. Upon bending of the substrate, a nanosized gap is formed between the electrodes, which can be mechanically adjusted and which is impressively stable on the order of hours, even at room temperature [13,14]. The layout of the technique is schematically presented in Fig. 1(b).

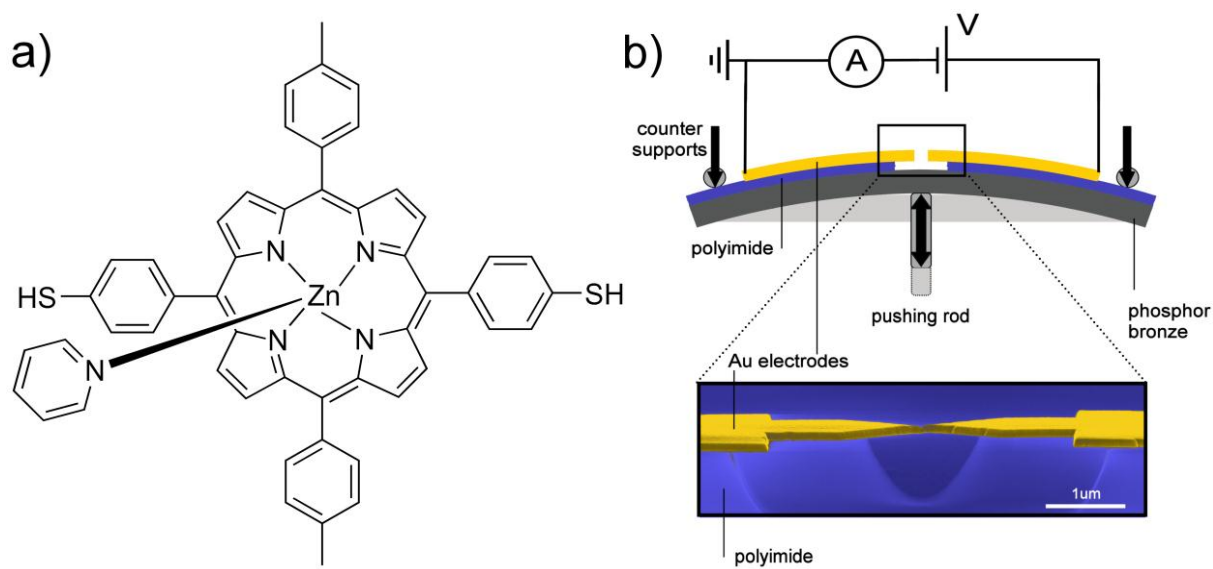

**Figure 1** Structural formula of ZnTPPdT-Pyr (b) Top: lay-out of the mechanically controllable break junction (MCBJ). Bottom: scanning electron micrograph of a MCBJ device (colorized for clarity). The scale bar shows that the suspended bridge is about 1 μm.

All experiments are performed in high vacuum (<$10^{-6}$ mbar). Prior to the experiments, [Zn$^{II}$(TPPdT)(Pyr)] (TPPdT stands for (5,15-di(p-thiolphenyl)-10,20-di(p-tolyl)porphyrin, Pyr for pyridine, see Fig. 1a for the structural formula, abbreviated as ZnTPPdT-Pyr in the following) is dissolved in dichloromethane (DCM) and deposited on the unbroken electrodes using self-assembly from solution. Two thiol groups on opposite sides of the molecule are used as anchoring groups. After deposition, the junctions are broken in vacuum at room temperature. The aforementioned stability of the electrodes allows us to characterize charge transport through ZnTPPdT-Pyr by performing two types of experiments. First, we measure at room temperature the low-bias conductance of the molecule as a function of electrodes stretching. Second, we perform spectroscopy of the molecular levels by measuring current-voltage characteristics, *I(V)s,* at fixed electrode spacings; This has been done both at room temperature and cryogenic temperature (6K).

## Results

To obtain the conductance value of the most probable contact geometry we repeatedly break and fuse the electrodes [15-17]) between conductances of $1·10^{-5}$ $G_0$ and 10 $G_0$, while measuring the current at a fixed bias voltage (100 mV). Each breaking event produces a `breaking trace´ of the conductance (plotted as $\log_{10}(G)$) versus the electrode displacement (d). Sets of 500 consecutive breaking traces from individual junctions are then binned in time and in electrode displacement. As we are interested in the breaking dynamics of the junctions beyond the point of rupture of the last monatomic gold contact (defined as d=0), only conductance values below one quantum unit $G_0=2e^2/h$ (the resistance of a single gold atom) are considered. The results are plotted as two-dimensional `trace histograms', in which areas of high counts represent the most typical breaking behavior of the molecular junction [18-19].

In Fig. 2 we show trace histograms as well as examples of individual breaking traces for a junction exposed to (a) the solvent DCM and (b) ZnTPPdT-Pyr. All measured curves are included, i.e., no data selection has been employed. We measured several samples with ZnTPPdT-Pyr molecules as well as DCM references. The features shown in Fig. 2a and 2b. are representative for all these measurements. In the junction that was exposed to the pure solvent without porphyrin molecules (Fig. 2a), the Au-bridge is stretched until a single-atom contact is formed, visible (only in the individual traces in the inset) as a plateau around the conductance quantum (G ~ $G_0$). Upon further stretching, the monatomic contact is broken and the conductance decreases sharply and abruptly to ~$10^{-3}$ $G_0$ due to relaxation of the electrode tips. Beyond this point, electron tunneling between the electrodes leads to a fast conductance decay with stretching (visible as the orange tail), as expected for tunneling through a single barrier.

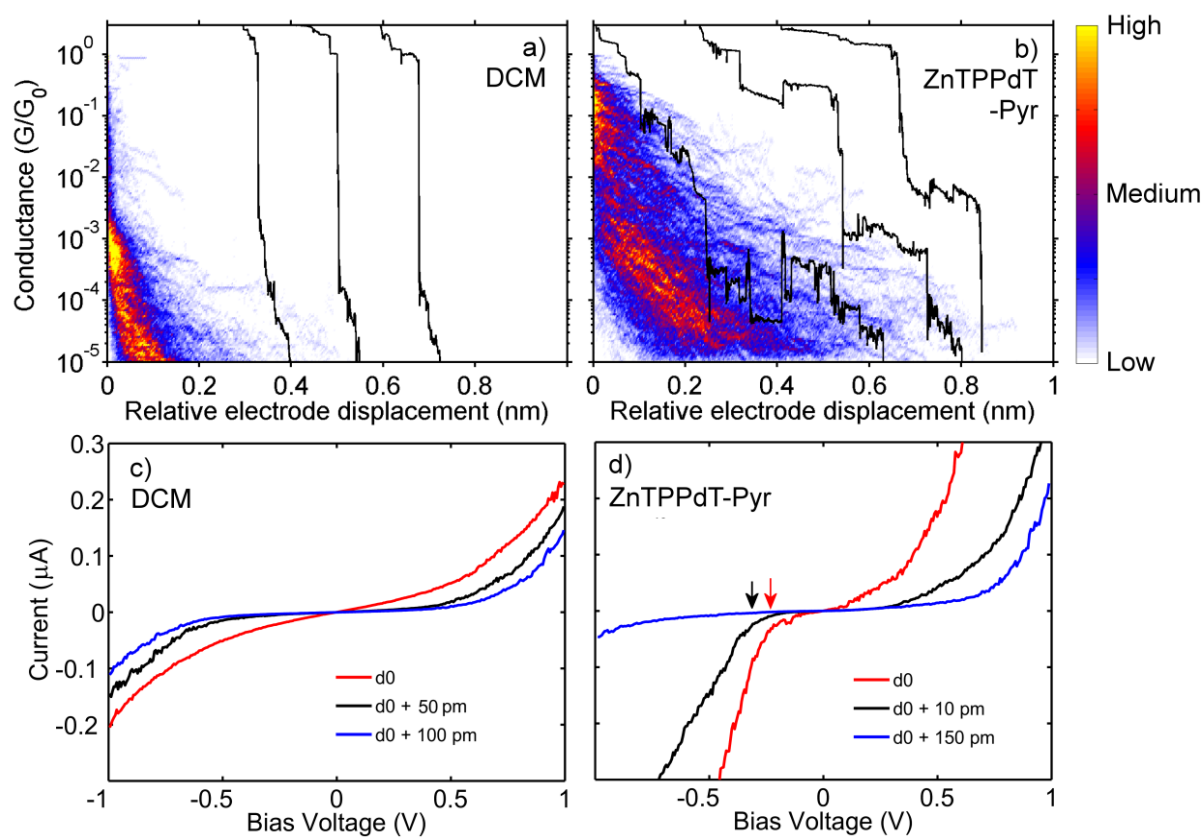

**Figure 2** Trace histograms constructed from 500 consecutive breaking traces taken at room temperature and 100mV bias for junctions exposed to (a) the solvent DCM only, (b) ZnTPPdT-Pyr. Regions of high counts represent the most probable breaking behavior of the contact. The black curves are examples of individual breaking traces (offset along the horizontal axis, d, for clarity). For the construction of the trace histograms the zero of the relative electrode displacement for each curve was set to the point where the conductance drops sharply below 1 $G_0$. (c) Current-voltage characteristics taken at various electrode spacings starting from the initial value $d_0$ of junctions exposed to the solvent DCM, and (d) ZnTPPdT-Pyr.

In contrast to this fast tunneling decay, introducing the porphyrin molecules by self-assembly on the junction leads to pronounced plateaus at different conductance values in the sub-$G_0$ regime. The observation of such plateaus in the breaking traces is commonly taken as a signature of the formation of a molecular junction [15-17]. Figure 2b shows that the plateaus can be horizontal or sloped. Some traces consist of a few plateaus at different conductance values. The representative breaking traces that are included in Fig. 2b display a set of such plateaus. In strong contrast to measurements on rod-like molecules, averaging over 500 traces does not lead to a narrow region of high counts in the trace histograms. Instead, two distinct regions with high counts are visible; a high-conductance region around $10^{-1}$ $G_0$, and a sloped low-conductance region ranging from $10^{-3}$ $G_0$ to $10^{-5}$ $G_0$. Although clear plateaus are observed in the single breaking traces, averaging over hundreds of traces washes out the molecular signature. Hence, a complementary method is required to study charge transport in more detail.

We have therefore measured current-voltage (*I(V)*) characteristics at fixed electrode spacing, in the $10^{-2}$ - $10^{-5}$ $G_0$ conductance region. In between the I(V)'s, the inter-electrode distance was gradually increased or decreased in steps of about 10 pm, without fusing the electrodes to form a metallic contact. In this way, changes in molecular junction configurations occurring as a function of electrode spacing can be accurately probed. I(V)'s taken at room temperature for a few settings of the electrodes spacing of junctions exposed to DCM and ZnTPPdT-Pyr are presented in

Fig. 2c and Fig. 2d, respectively. For each series, all presented I(V)'s are taken from the same breaking sequence.

I(V)'s of a junction exposed to DCM (Fig. 2c) exhibit the characteristic single-barrier tunneling shape and show the expected current decrease upon increasing the electrode spacing. In contrast, I(V) characteristics on the ZnTPPdT-Pyr junction show a sharper current onset, marked by arrows in figure Fig. 2d. This observation may be viewed as a molecular fingerprint as they correspond to the onset of resonant transport through an energy level of the molecule (either vibrational or electronic). Interestingly, the current onset strongly depends on the inter-electrode distance. At $d_0$ it is located around -250 mV. After a step of about 10 pm in the electrode distance, the onset shifted to around -350mV. Increasing the inter-electrode distance with an additional 140 pm, shifts the onset at negative bias to a location outside the bias window. Note furthermore the asymmetry in the curves in Fig. 2d, which increases as they electrode moves further apart (e.g. the blue curve in this figure). For the three I(V)'s we have also determined the conductance at the same bias voltage as used to construct the trace histograms, i.e., at 100mV. For the red, black and blue I(V) curve we obtain conductance values of $2.0 \cdot 10^{-3}$, $1.6 \cdot 10^{-4}$ and $1.6 \cdot 10^{-4}$ $G_0$ respectively. Interestingly, small changes in electrode distance (~10 pm) can induce big changes in the shape of the I(V) characteristics and the low-bias conductance (compare e.g. the red and black curves). Opening the junction further (black and blue curves) results in an equal conductance value at 100 mV, but different I(V) shapes.

Spectroscopic features become more pronounced at low temperature as the junction stability increases, and both the thermal noise and thermal broadening decrease. We therefore cooled down the junctions to cryogenic temperature (6 K) while keeping the zero-bias conductance at a fixed value (around $1\cdot 10^{-4}$ $G_0$) with a feedback loop. In Fig. 3 (a) (b) we present low temperature I(V) characteristics of junctions exposed to (a) DCM and (b) ZnTPPdT-Pyr solution, for different electrode spacings. IV's of the junction exposed to DCM show the characteristic tunneling shape, without any molecular signature, as was also found at room temperature. A notable difference, however, is the significant reduction of the noise.

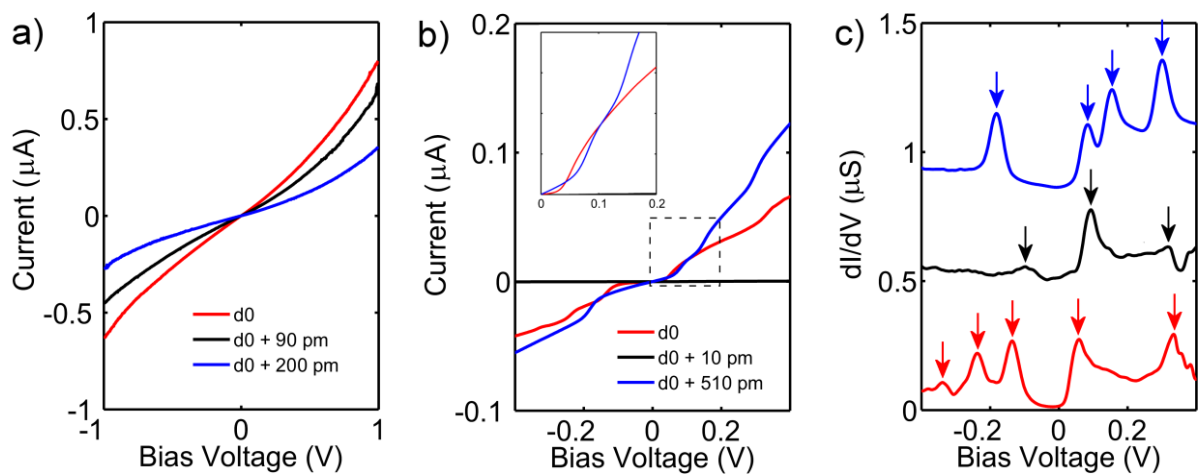

**Figure 3** I(V) characteristics of junctions exposed to (a) DCM and (b) ZnTPPdT-Pyr. The DCM sample clearly shows vacuum tunneling behavior. The molecular sample exhibits Coulomb blockade and steps. (c) dI/dV of a junction exposed to a ZnTPPdT-Pyr solution, offset vertically for clarity. Resonances correspond to electronic or vibrational energy levels of the molecular junction. Note, for the black line the dI/dV has been multiplied by 100.

The IV's of the ZnTPPdT-Pyr containing junction now show sharp step-like features, which are more pronounced than that in Fig. 2d. We numerically determined the differential conductance (dI/dV) as displayed in Fig. 3c. In the dI/dV's, the step-like features are visible as resonances, which have been marked in the figure with arrows of the corresponding color. For clarity, the dI/dV's have been offset vertically, and the dI/dV represented by the black curve magnified 100 times. The origin of those resonances can be electronic or vibrational [20-22]. Independently on their origin, their position reveals the alignment of the corresponding energy level with respect to the Fermi Energy of the electrodes [23]. For a distance of $d_0$ (red curve), five pronounced resonances are present, located at -339mV, -283mV, -153mV, 58mV and 334mV. For the conductance at 100mV we obtain a value of $2.1 \cdot 10^{-3}$ $G_0$. Increasing the distance by 10 pm (black curve) drastically changes the molecular energy spectrum, with one distinct resonance at 94mV, and two fainter peaks around -99mV and 319 mV. Here, the conductance at 100 mV is $1.2 \cdot 10^{-5}$ $G_0$. Increasing the distance with an additional 500 pm (blue curve) again leads to changes in the molecular energy spectrum; four pronounced resonances are now located at -238mV, -136mV, 58mV and 334mV. For the conductance we again obtain a value of $2.1 \cdot 10^{-3}$ $G_0$.

## Discussion

Comparing first the red and black curve in Fig. 3c, we see that within an electrode displacement of 10 pm, the number of energy levels involved in the electronic transport as well as their energy has drastically changed. A major jump of two orders of magnitude in the low-bias conductance is observed as well. This suggests an abrupt change in the molecule-electrode interaction, presumably caused by a change in molecular configuration. A similar change in molecular configuration was also observed in the room temperature IV's for the red and black curve; the onset for current shifted by -100mV and the conductance dropped one order of magnitude within 10 pm. These observations support the conclusion drawn from the trace histogram measurements: The molecule can adopt different stable configurations, leading to plateaus at different conductance values in the breaking traces. Comparing the red and blue curve in Fig. 3c, which are taken are at a separation of

510 pm, we see that their molecular energy spectrum strongly differ, but that their low-bias conductance is similar (Fig 3b, inset). Similar behavior is also observed at room temperature (Fig. 2d). This suggests that different stable junction configurations with very different spectroscopic signatures can exhibit the same low-bias conductance.

For most of the low-bias break junction measurements on rod-like molecules it is assumed that repetitive fusing and breaking of the molecular junction provides the most probable conductance value [15-17]. Multiple conductance peaks are often attributed to the formation of multiple molecular bridges connected in parallel [15,24]. The strength of the metal-molecule chemical bond is considered to play a central role in determining the single molecule conductance values. Our results on the Zn porphyrin molecule with a pyridine axial group show that different conductance values can also result from stretching or fusing a molecular junction. As considerable changes in the conductance values and spectra already occur for a displacement as small as 10 pm, we conclude that neither the chemical molecule-electrode bond nor the electrode configuration itself can be held responsible. More likely, varying the electrode distance changes the molecular configuration, which in turn leads to abrupt changes in the molecule-electrode interaction. Our findings also show that I(V) characteristics taken at different electrode spacings can exhibit very distinct spectroscopic features but a similar low bias conductance. This indicates that different junction geometries can lead to similar conductance values in the trace histograms. Therefore, as changes in the molecular junction conformation are not always reflected in the low-bias trace histograms, supporting high bias IV-characteristics are essential for the interpretation of such histograms.

# Conclusion

In summary, we investigated charge transport in ZnTPPdT-Pyr molecular junctions using the lithographic MCBJ technique. We combined low-bias statistical measurements with spectroscopy measurements of the molecular levels using I(V) characteristics. This unique combination allows us to probe different junction configurations and monitor changes in the molecular level alignment upon fusing or breaking of a molecular junction. Both methods show that multiple stable single–molecule junction configurations can be obtained by stretching or fusing the junction. In addition we demonstrate that different ZnTPPdT-Pyr junction configurations can lead to different spectroscopic features for similar low-bias conductance values. Thus, I(V) spectroscopy measurements can provide additional information compared to statistical low-bias conductance histograms, enabling a more in-depth characterization of the charge transport through a single molecule.

# References

[1] Xiao, X.; Xu, B.; Tao, N., *Journal of the American Chemical Society*, **2004,** 126, 5370-5371

[2] Li, X.; He, J.; Hihath, J.; Xu, B.; Lindsay, S.M.; Tao, N., *Journal of the American Chemical Society*, **2006**, 128, 2135-2141

[3] Venkataraman, L.; Klare, J.E.; Nuckolls, C.; Hybertsen, M.S.; Steigerwald, M.L., *Nature*, **2006**, 442, 904-907

[4] Mishchenko, A.; Vonlanthen, D., Meded, V.; Bürkle, M.; Li, C.; Pobelov, I.V.; Bagrets, A.; Viljas, J.K.; Pauly, F.; Evers, F.; Mayor, M.; Wandlowski, T., *Nano Letters*, **2010**, 10, 156–163

[5] Park, Y.S.; Whalley, A.C.; Kamenetska, M.; Steigerwald, M.L.; Hybertsen, M.S.; Nuckolls, C.; Venkataraman, L., *Journal of the American Chemical Society*, **2007**, 129, 15768-15769

[6] Chen, F.; Li, X.; Hihath, J.; Huang, Z.; Tao, N., *Journal of the American Chemical Society*, **2006**, 128, 15874-15881

[7] Qiu, X. H.; Nazin, G.V.; Ho, W., *2004, Physical Review Letters*, 93, 196806

[8] Brede, J.; Linares, M.; Kuck, S.; Schwöbel, J.; Scarfato, A.; Chang, S.-H.; Hoffmann, G.; Wiesendanger, R.; Lensen, R.; Kouwer, P. H. J.; Hoogboom, J.; Rowan, A. E.; Bröring, M.; Funk, M.; Stafström, S.; Zerbetto, F.; Lazzaroni, R. *Nanotechnology* **2009**, *20*, 275602.

[9] Perrin, M.L.; Prins, F.; Martin, C.A.; Shaikh, A.J.; Eelkema, R.; van Esch, J.H.; Briza, T.; Kaplanek, R.; Kral, V.; van Ruitenbeek, J.M.; van der Zant, H.S.J.; Dulić, D., *Angewandte Chemie International Edition*, **2011**, *in press*.

[10] van Ruitenbeek, J.M.; Alvarez, A.; Piñeyro, I.; Grahmann, C.; Joyez, P.; Devoret, M.H.; Esteve, D.; Urbina, C., **1996**, *Review of Scientific Instruments*, 67, 108-111.

[11] Rubio, G.; Agraıt, N.; Vieira, S., **1996**, *Physical Review Letters*, 76, 2302 -2305.

[12] Scheer, E.; Agraït, N.; Cuevas, J.C.; Yeyati, A.L.; Ludoph, B.; Martín-Rodero, A.; Bollinger, G.R.; van Ruitenbeek, J.M.; Urbina, C., **1998**, *Nature*, 394, 154-157.

[13] Martin, C.A.; Ding, D.; van der Zant; van Ruitenbeek, J.M., **2008**, *New Journal of Physics*, 10, 065008

[14] Dulic, D.; Pump, F.; Campidelli, S.; Lavie, P.; Cuniberti, G.; Filoramo, A., **2009**, *Angewandte Chemie International Edition*, 48, 8273 –8276.

[15] Xu, B.Q.; Tao, N.J., **2003**, *Science*, 301, 1221-1223.

[16] Gonzalez, M.T; Wu, S.; Huber, R.; van der Molen, S.J.; Schönenberger, C.; Calame, M., **2006**, *Nano Letters*, 6, 2238-2242.

[17] Venkataraman, L.; Klare, J.E.; Tam, I.W.; Nuckolls, C.; Hybertsen, M.S.; Steigerwald, M.L., **2006**, *Nano Letters*, 6, 458-462.

[18] Martin, C.A.; Ding, D.; Sørensen, J.K.; Bjørnholm, T.; van Ruitenbeek, J.M.; van der Zant, H.S.J, **2008**, *Journal of the American Chemical Society*, 130,13198–13199.

[19] Kamenetska, M.; Koentopp, M.; Whalley, A.C.; Park, Y.S.; Steigerwald, M.L.; Nuckolls, C.; Hybertsen, M.S.; Venkataraman, L., **2009**, *Physical Review Letters*, 102, 126803

[20] Osorio, E. A.; O'Neill, K.; Stuhr-Hansen, N.; Nielsen, O.F.; Bjørnholm, T.; van der Zant, H. S. J., **2007**, *Advanced Materials*, 19, 281–285.

[21] Reichert, J.; Ochs, R.; Beckmann, D.; Weber, H.B.; Mayor, M.; v. Löhneysen, H.V., **2002**, *Physical Review Letters*, 88, 176804.

[22] Kubatkin, S.; Danilov, A.; Hjort, M.; Cornil, J.; Brédas, J.-L.; Stuhr-Hansen, N.; Hedegård, P.; Bjørnholm, T, **2003**, *Nature*, 425, 698-701.

[23] Thijssen, J.M.; van der Zant, H.S.J., **2008**, *Physica Status Solidi (b), 245,* 1–16.

[24] Gonzalez, M.T; Brunner, J.; Huber, R.; Wu, S.; Schönenberger, C.; Calame, M., **2008**, *New Journal of Physics*, 10, 065018.